# Entangling power and local invariants of two-qubit gates

S. Balakrishnan   and   R. Sankaranarayanan

*Department of Physics, National Institute of Technology, Tiruchirappalli 620015, India*

We show a simple relation connecting entangling power and local invariants of two-qubit gates. From the relation, a general condition under which gates have same entangling power is derived. The relation also helps in finding the lower bound of entangling power for perfect entanglers, from which the classification of gates as perfect and non perfect entanglers is obtained in terms of local invariants.

Entanglement, a nonlocal property of a quantum state, is regarded as a resource for realizing various fascinating features such as teleportation, quantum cryptography and quantum computation [1, 2]. On one side, much work has been carried out to understand and exploit the entanglement for various information processing. On the other side, attention has been given to quantum operations (gates) as they are responsible for creating entanglement when acting on a state.

Since two-qubit gates are capable of producing entanglement, it is of vital importance to understand their entangling characterization. One such useful tool is the entangling power of an operator $e_P(U)$ which quantifies the average entanglement produced [3]. Another tool to characterize the nonlocal attributes of a two-qubit gate is local invariants, namely $G_1$ and $G_2$ (first introduced in Ref. [4]) such that gates differing only by local operations possess same invariants. Furthermore, nonlocal two-qubit gates form an irreducible geometry of tetrahedron known as Weyl chamber. Of all the gates, exactly half of them are *perfect entanglers* (operators capable of producing maximally entangled state from some input product state) and they form a polyhedron within the Weyl chamber [5].

It is known that gates differing only by local operations possess the same entangling power. Similarly, gates which are inverse to each other possess the same entangling power. For instance, $SWAP^{\alpha}$ and $SWAP^{-\alpha}$ assume the same entangling power as they are inverse to each other. From our earlier study on the geometrical edges of two-qubit gates [6], it was found that gates which do not belong to the preceding category also possess same entangling power. For example, entangling power of the gates lying in the polyhedron edges *QP*, *MN* and *PN* are identical [6]. Motivated by this fact, here we investigate the entangling power of two-qubit gates in detail. In this brief report, we establish a simple relation between the entangling power and local invariants. It is shown that if the $|G_1|$ of two gates is the same, they possess the same entangling power. The relation also facilitates in showing that the minimum entangling power of perfect entanglers is possessed by the three edges of the polyhedron mentioned earlier. Furthermore, we find the conditions for the perfect entanglers in terms of local invariants, which are useful for the classification of two-qubit gates as perfect and non perfect entanglers.

Let us consider a general two-qubit gate $U$ [7]:

$$U = \begin{pmatrix} e^{-\frac{ic_3}{2}}c^- & 0 & 0 & -ie^{-\frac{ic_3}{2}}s^- \\ 0 & e^{\frac{ic_3}{2}}c^+ & -ie^{\frac{ic_3}{2}}s^+ & 0 \\ 0 & -ie^{\frac{ic_3}{2}}s^+ & e^{\frac{ic_3}{2}}c^+ & 0 \\ -ie^{-\frac{ic_3}{2}}s^- & 0 & 0 & e^{-\frac{ic_3}{2}}c^- \end{pmatrix} \quad (1)$$

where $c^\pm = \cos[(c_1 \pm c_2)/2]$, $s^\pm = \sin[(c_1 \pm c_2)/2]$ and $[c_1, c_2, c_3]$ is the geometrical point of a two-qubit gate [4, 5]. We note that the geometrical representation of two-qubit gates (Weyl chamber) is described by $c_1 \geq c_2 \geq c_3 \geq 0$. The entangling capability of a unitary quantum gate $U$ can be quantified by the entangling power which is defined as [3, 8]

$$e_P(U) = \overline{[E(U|\psi_1\rangle \otimes |\psi_2\rangle)]}_{|\psi_1\rangle \otimes |\psi_2\rangle} \quad (2)$$

where the overbar denotes the average over all product states distributed uniformly in the state space. In the preceding formula $E$ is the linear entropy of entanglement measure defined as

$$E(|\psi\rangle_{AB}) = 1 - tr(\rho_{A(B)}^2) \quad (3)$$

where $\rho_{A(B)} = tr_{B(A)}(|\psi\rangle_{AB}\langle\psi|)$ is the reduced density matrix of system *A(B)*.

The expression to calculate the entangling power of a two-qubit gate $U$ is [3, 9]:

$$e_P(U) = \frac{5}{9} - \frac{1}{36}\{\langle U^{\otimes 2}, T_{1,3}U^{\otimes 2}T_{1,3}\rangle + \langle (SWAP \times U)^{\otimes 2}, T_{1,3}(SWAP \times U)^{\otimes 2}T_{1,3}\rangle\} \quad (4)$$

where $\langle A, B\rangle = tr(A^\dagger B)$ is referred to as Hilbert – Schmidt scalar product and $T_{1,3}$ is the transposition operator defined as $T_{1,3}|a,b,c,d\rangle = |c,b,a,d\rangle$ on a four-qubit system. In what follows we use the definitions: $A = U^{\otimes 2}$, $S = SWAP^{\otimes 2}$, $B = (SWAP \times U)^{\otimes 2}$ and $T = T_{1,3}$. Exploiting the property of tensor products [10]: $(A_1 A_2) \otimes (B_1 B_2) = (A_1 \otimes B_1)(A_2 \otimes B_2)$ we can write $B = SA$. With this, we have $\langle B, TBT \rangle = tr(A^\dagger S^\dagger TSAT)$ and hence the entangling power can be rewritten as

$$e_P(U) = \frac{5}{9} - \frac{1}{36}[tr(A^\dagger TAT) + tr(A^\dagger S^\dagger TSAT)] \qquad (5)$$

Using the fact that $tr(A) + tr(B) = tr(A + B)$, we write entangling power as

$$e_P(U) = \frac{5}{9} - \frac{1}{36}[tr(A^\dagger RAT)] \qquad (6)$$

where $R = T + S^\dagger TS$. Substituting Eq. (1) in the preceding expression, after some simplifications, the entangling power can be rewritten as

$$e_P(U) = \frac{2}{9}[1 - |G_1|] \qquad (7)$$

where

$$|G_1| = cos^2 c_1 cos^2 c_2 cos^2 c_3 + sin^2 c_1 sin^2 c_2 sin^2 c_3. \qquad (8)$$

Thus we obtain a simple relation between the entangling power $e_P$ and local invariant $G_1$ of a two-qubit gate. The relation also implies that gates having the same $|G_1|$ must necessarily possess the same $e_P$. Since the invariant $G_1$ for a gate and its inverse are complex conjugate to each other, both the gates will have same $e_P$. Since $0 \leq |G_1| \leq 1$, it is evident that $0 \leq e_P \leq 2/9$. Here, we note that Eq. (7) can also be rewritten as [7]

$$e_P(U) = \frac{1}{18}[3 - (cos2c_1 cos2c_2 + cos2c_2 cos2c_3 + cos2c_3 cos2c_1)]. \qquad (9)$$

In our earlier study on the geometrical edges of polyhedron, it was shown that $e_P = 1/6$ for the edges $QP$, $MN$ and $PN$ [6]. In terms of Eq. (7), this result is understandable as $|G_1| = 1/4$ for all these edges. We also note that the identical parameter dependence of $e_P$ for the other edges of polyhedron: $LQ$, $LN$ and $A_2P$, is also reflected through their $|G_1|$ [6]. Furthermore, Eq. (7) is also useful to identify the gates with maximum and minimum $e_P$. If $e_P = 2/9$, $|G_1| = 0$ which is possible only for $[\pi/2, \varphi, 0]$ where $0 \leq \varphi \leq \pi/2$. These gates correspond to the well known family of special perfect entanglers (SPE) [7]. If $e_P = 0$, $|G_1| = 1$ which is possible only for (i) $[0,0,0]$, a local gate and (ii) $[\pi/2, \pi/2, \pi/2]$, SWAP gate.

A two-qubit gate is called a perfect entangler (PE) if it produces a maximally entangled state for some input product state [11]. Considering the symmetry in the maximal entanglement production by the gates, we confine our attention to one half of the Weyl chamber: $\pi/2 \geq c_1 \geq c_2 \geq c_3 \geq 0$. If the geometrical points are such that

$$\text{(A) } c_1 + c_2 \geq \pi/2 \quad \text{and} \quad \text{(B) } c_2 + c_3 \leq \pi/2 \qquad (10)$$

then the corresponding gate is a perfect entangler [11].

Having known that SPE possess the maximum $e_P$, here we exploit Eq. (7) to identify PEs which possess minimum $e_P$. In other words, we find PEs which possess maximum $|G_1|$. Let us rewrite the first term of Eq. (8) as

$$cos^2 c_1 cos^2 c_2 cos^2 c_3 = \{[cos(c_1 + c_2) + cos(c_1 - c_2)]^2 cos^2 c_3\}/4. \tag{11}$$

Imposing the condition (A) implies that $-1 \leq cos\,(c_1 + c_2) \leq 0$ and $0 \leq cos\,(c_1 - c_2) \leq 1$. Then $|G_1|$ has the maximum value of 1/4 only for $c_1 = c_2 = \pi/4$, for which the condition (B) becomes $0 \leq c_3 \leq \pi/4$. In other words, the edge $QP\,[\pi/4, \pi/4, \eta]$ with $0 \leq \eta \leq \pi/4$ is such that $|G_1| = 1/4$ and hence $e_P = 1/6$ [6]. It is worth recollecting that if $[c_1, c_2, c_3]$ is a perfect entangler then $[\pi - c_1, c_2, c_3]$ is also a perfect entangler. Since the edge $QP\,[\pi/4, \pi/4, \eta]$ is a PE, the edge $MN\,[3\pi/4, \pi/4, \eta]$ is also PE with $e_P = 1/6$. In a similar way, the second term of Eq. (8) is rewritten as

$$sin^2 c_1 sin^2 c_2 sin^2 c_3 = \{sin^2 c_1 [cos(c_2 - c_3) - cos(c_2 + c_3)]^2\}/4. \tag{12}$$

Imposing the condition (B) implies that $0 \leq cos\,(c_2 + c_3) \leq 1$ and $0 \leq cos\,(c_2 - c_3) \leq 1$. Then $|G_1|$ has the maximum value of 1/4 only for $c_2 = c_3 = \pi/4$, for which the condition (A) becomes $\pi/4 \leq c_1 \leq \pi/2$. In other words, the gates $\left[\frac{\pi}{4} + \theta, \frac{\pi}{4}, \frac{\pi}{4}\right]$ with $0 \leq \theta \leq \pi/4$ are such that $|G_1| = 1/4$ and $e_P = 1/6$. Since the gates $\left[\frac{\pi}{4} + \theta, \frac{\pi}{4}, \frac{\pi}{4}\right]$ are PEs, the gates $\left[\frac{3\pi}{4} - \theta, \frac{\pi}{4}, \frac{\pi}{4}\right]$ are also PEs with $e_P = 1/6$. Alternatively, the edge $PN\,\left[\frac{\pi}{4} + \eta, \frac{\pi}{4}, \frac{\pi}{4}\right]$ with $0 \leq \eta \leq \pi/2$ is a PE having $e_P = 1/6$. From the proceding analysis, it is clear for the PE that $1/6 \leq e_P \leq 2/9$. In this range, SPE possess the maximum and the polyhedron edges $QP$, $MN$ and $PN$ possess the minimum. In terms of local invariant $G_1$ the preceding inequality reads as $0 \leq |G_1| \leq 1/4$.

Having found the range of $G_1$, the following theorem identifies the range of local invariant $G_2$ for PEs.

*Theorem 1.* PEs are such that $-1 \leq G_2 \leq 1$.

*Proof.* The expression for $G_2$ is as given in Ref. [5]:

$$G_2 = 4cos^2 c_1 cos^2 c_2 cos^2 c_3 - 4sin^2 c_1 sin^2 c_2 sin^2 c_3 - cos 2c_1 cos 2c_2 cos 2c_3 \tag{13}$$

or [7]

$$G_2 = cos\,2c_1 + cos\,2c_2 + cos\,2c_3. \tag{14}$$

The preceding expression is rewritten as

$$G_2 = 2cos(c_1 + c_2)\,cos(c_1 - c_2) + cos\,2c_3. \tag{15}$$

On imposing the condition (A), we have $-1 \leq \cos(c_1 + c_2) \leq 0$ and $0 \leq \cos(c_1 - c_2) \leq 1$. Then $G_2$ has the maximum value of 1 for $c_1 = \frac{\pi}{2} - \theta$, $c_2 = \theta$ and $c_3 = 0$. In other words, the edge $LQ\left[\frac{\pi}{2} - \theta, \theta, 0\right]$ of Weyl chamber with $0 \leq \theta \leq \frac{\pi}{4}$ is such that $G_2 = 1$ [6]. Similarly, $G_2$ takes the minimum value of -1 for $c_1 = c_2 = \pi/2$ and $c_3 = 0$, which corresponds to DCNOT. Hence, the proof is completed.

From the earlier analysis on $G_1$, we observe that *all* the PEs lie within the range $0 \leq |G_1| \leq 1/4$. It is worth mentioning that non PEs are also found within this range, for example, some controlled unitary gates [6]. In order to classify the gates based on the local invariants, we prove the following theorem.

*Theorem 2.* Non PEs lie within the range $0 \leq |G_1| \leq 1/4$, do not satisfy $-1 \leq G_2 \leq 1$.

*Proof.* Consider all the non PEs within the range $0 \leq |G_1| \leq 1/4$, for which the condition (A) or (B) must be violated. Violation of both (A) and (B) amounts to the violation of Weyl chamber condition: $\pi/2 \geq c_1 \geq c_2 \geq c_3 \geq 0$. First let us assume that (A) is violated, i.e., $c_1 + c_2 < \pi/2$. It implies that $0 < \cos(c_1 + c_2) \leq 1$ and $0 < \cos(c_1 - c_2) \leq 1$, and $G_2$ takes the maximum value of 3 for $c_1 = c_2 = c_3 = 0$ (local gate). In order to find the minimum value of $G_2$, we take $c_1 + c_2 = \frac{\pi}{2} - \epsilon$ and $c_1 - c_2 = \frac{\pi}{2} - \delta$ where $0 < \epsilon < \delta \ll 1$. Then, the minimum value of $G_2$ is $1 + 3\epsilon\delta$ for $c_1 = (\pi - \epsilon - \delta)/2$, $c_2 = c_3 = (\delta - \epsilon)/2$. Secondly, let us consider that (B) is violated, i.e., $c_2 + c_3 > \frac{\pi}{2}$. Rewriting Eq. (14) as

$$G_2 = \cos 2c_1 + 2\cos(c_2 + c_3)\cos(c_2 - c_3), \quad (16)$$

we have $-1 \leq \cos(c_2 + c_3) < 0$ and $0 < \cos(c_2 - c_3) \leq 1$. Then the minimum value of $G_2$ is -3 for $c_1 = c_2 = c_3 = \frac{\pi}{2}$ (SWAP gate). For the maximum value of $G_2$, we take $c_2 + c_3 = \frac{\pi}{2} + \epsilon$ and $c_2 - c_3 = \frac{\pi}{2} - \delta$. With this the maximum value of $G_2$ is $-(1 + 3\epsilon\delta)$ for $c_1 = c_2 = (\pi + \epsilon - \delta)/2$, $c_3 = (\epsilon + \delta)/2$. Hence, non PEs which violate (A) or (B) do not fall in the range $-1 \leq G_2 \leq 1$ and the proof is completed.

From the theorem 2, we conclude that PEs satisfy

$$(C)\ 0 \leq |G_1| \leq 1/4 \quad \text{and} \quad (D) -1 \leq G_2 \leq 1, \quad (17)$$

and the gates that do not satisfy both these conditions are non PEs. It is easy to recognize that Eq. (17) and Eq. (10) are equivalent. Thus the local invariants associated to a gate are found to useful for the classification as perfect and non perfect entanglers. It is worth emphasizing that Eq. (17) involves two parameters namely, $|G_1|$ and $G_2$, while Eq. (10) involves three geometrical parameters.

In this work, we have shown a simple relation between entangling power $e_p$ and local invariant $G_1$ of a two-qubit gate. The relation implies that the gates with same $|G_1|$ possess same entangling power. Thus the local invariant $G_1$ of a gate also signifies the average entanglement produced. Gates differing only by local operations have same $G_1$ and hence $e_p$. Since the invariant $G_1$ for a gate and its inverse is complex conjugate to each other, both the gates will have same $e_P$. It is identified that three geometrical edges of polyhedron, namely QP, MN and PN are such that $|G_1| = 1/4$ and hence $e_p = 1/6$. It is shown for perfect entanglers that $1/6 \leq e_P \leq 2/9$ or $0 \leq |G_1| \leq 1/4$, such that $e_P = 1/6$ for the preceding three edges and $e_P = 2/9$ for special perfect entanglers.

Furthermore, the local invariant $G_2$ for the perfect entanglers are such that $-1 \leq G_2 \leq 1$. From the obtained range of local invariants $G_1$ and $G_2$, it is shown that the invariants are also useful to classify the two-qubit gates as perfect and non perfect entanglers. It is worth noting that the obtained classification based on local invariants does not require the geometrical point of a gate.

S. B acknowledges financial support from Council of Scientific and Industrial Research, India. S. B also acknowledges Pankaj Agarwal and Arun K Pati, Institute of Physics, Bhubaneswar, India for their encouragement in working on this problem.